# Effect of LaAlO$_3$ twin-domain topology on local dc and microwave properties of cuprate films

A. P. ZHURAVEL

*B. Verkin Institute for Low Temperature Physics & Engineering, NAS of Ukraine*
*Kharkov 61103, Ukraine*

STEVEN M. ANLAGE

*Department of Physics, Center for Nanophysics and Advanced Materials*
*University of Maryland, College Park, Maryland 20742-4111, USA*

STEPHEN K. REMILLARD

*Department of Physics, Hope College, 27 Graves Place, Holland, Michigan 49423, USA*

A.V. LUKASHENKO AND A. V. USTINOV

*Physikalisches Institut, Karlsruhe Institute of Technology, D-76128 Karlsruhe, Germany*
*DFG-Center for Functional Nanostructures (CFN), D-76128 Karlsruhe, Germany*

*Abstract* — Different imaging modes of low temperature laser scanning microscopy (LTLSM) have been applied to probe local optical and superconducting properties, as well as the spatial variations in thermoelectric and electronic (both dc and rf) transport, in a YBa$_2$Cu$_3$O$_{6.95}$/LaAlO$_3$ (YBCO/LAO) superconducting microstrip resonator with micron-range resolution. Additionally, the local sources of microwave nonlinearity (NL) were mapped in two-dimensions simultaneously by using the LTLSM in two-tone rf intermodulation distortion contrast mode as a function of $(x, y)$ position of the laser beam perturbation on the sample. The influence of the direction of individual twin-domain YBCO blocks on its NL properties was analyzed in detail. The result shows the direct spatial correlation between NL microwave and dc electronic transport properties of the YBCO film that are imposed by the underlying twin-domain topology of the LAO substrate. In these circumstances, the scale of local NL current densities $J_{IM}(x, y)$ in different areas of the YBCO microstrip quantitatively coincide with the scale of local critical current densities $J_c(x, y)$ measured at the same positions.

## I. INTRODUCTION

In the last few years, there has been progress in the application of high-temperature superconducting (HTS) materials in passive microwave electronics including planar filters for mobile telephony. The technology is based on epitaxial growth of optimally doped copper oxide films YBa$_2$Cu$_3$O$_{6.95}$ (YBCO) on single crystal (100) lanthanum aluminate LaAlO$_3$ (LAO) substrates.[1] Thin ($d < \lambda_L$ ~200 nm at 4.2 K, where $\lambda_L$ is the London penetration depth) epitaxial YBCO films, with critical temperature of superconducting transition, $T_c \approx 92$ K above the boiling point of liquid nitrogen ($T_{LN2} \approx 77.3$ K), demonstrate very high critical current density, $J_c(T_B \approx 77$ K$) >(1–2)$ MA cm$^{-2}$ (where $T_B$ is the sample temperature), and low microwave surface resistance, R$_S(T_B$~77 K, $f = 10$ GHz$) < 0.5\ m\Omega$.[2] Thick ($d > 2\ \lambda_L$) YBCO coatings, in spite of their mainly polycrystalline microstructure, show similar characteristics. Both thick and thin planar HTS structures are well suited for manufacturing of high-$Q$ resonators with small microwave losses. In addition, crystal-lattice dimensions and the thermal expansion coefficients matched better than 2% at the

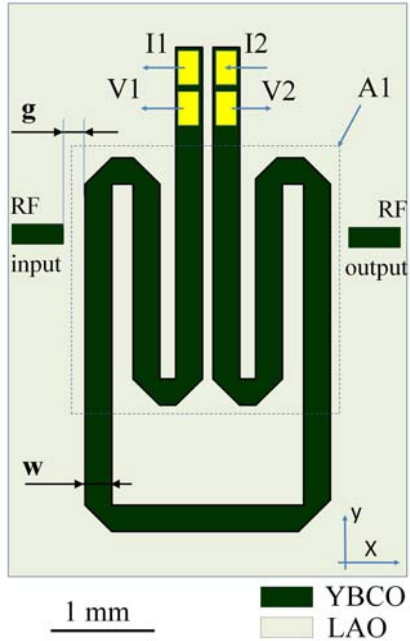

FIG. 1. Details of the HTS resonator geometry. dc bias is applied through current leads I1 and I2 while voltage leads V1 and V2 are used as potential probes to measure global IVCs as well as LSM photoresponse in dc resistive and unbiased TEI modes. Two capacitive gaps ($g$ =200 $\mu$m) separate the YBCO structure of the resonator from input and output rf electrodes used for microwave characterization of integral and local rf properties. The approximate area A1 of the LSM images in Fig. 3 is outlined by the dashed frame.

LAO/YBCO interface at the deposition temperature (650–900 °C). However, a structural phase transition $Pm - R\overline{3}c$ of LAO from rhombohedral at room temperature to the cubic perovskite aristotype structure at $T \sim 534$ °C activates substrate twinning [3, 4] that can affect microwave and superconducting properties of the YBCO film. This is due to the small coherence lengths of HTS materials that can be comparable to the nanoscale substrate surface roughness. Additionally, the YBCO material itself also is characterized by the formation of twin boundaries at $T_B \sim 400$ °C due to slight rearrangements of the atoms when crossing the tetragonal to orthorhombic structural transition.[5, 6] It was found that the superconducting order parameter [7] as well as critical current density [8] of YBCO is suppressed by both kinds of twin boundaries at low temperatures (LTs), while the twins attract or pin magnetic vortices. [9, 10] These twin boundaries cause an anisotropic dc flux guided flow of the Abrikosov vortices under the action of the Lorentz force, and dc Abrikosov vortices move preferentially along grain boundaries in YBCO including those formed between twin domains. [8, 11]

In this case, local electronic and magnetic transport properties of the HTS material are strongly dependent on local microstructure, which can be controlled by the growth conditions. [12] Such inhomogeneous microstructure leads to the spatial variation in $J_c$ and magnetic penetration depth, modifying the linear and nonlinear (NL) electrodynamics of the superconductor. The role of the LAO twinning and influence of the formed twin structures on dc and high-frequency (including rf) transport properties of thin HTS films were intensively discussed (see, for example, Refs. [9], [13], and [14]). As far as we know, such studies have not been conducted for thick HTS films. The problem here is that the nanoscale roughness of LAO substrates is small enough (compared to the film thickness) to not cause significant variations in $J_c$ across the polycrystalline HTS film. Furthermore, the microscopic structure of the film is not visually well defined. Both AFM and optical microscopy methods can provide only surface analysis and do not give useful information on the internal microstructure. And finally, correlation of microstructure with electronic properties of HTS films is also nontrivial due to the spatial averaging of most measurement methods. Thus, it is necessary to develop alternative spatially resolved methods that can directly determine the correlation between microstructure of both LAO and YBCO and transport dc and rf properties of the superconductor. Here we propose to use different imaging modes of the LT laser scanning microscopy (LTLSM) technique to study such kinds of problems. The initial subject of the present study is to understand the origin of microwave NLs in HTS materials, and to relate them to the physical microstructure of the material.

## II. SAMPLES

The experiments were carried out on a set of meandering microstrip HTS resonators designed to have the resonant frequency, $f_0$, at either a commercial PCS wireless communication band at 1.85 GHz or at a commercial cellular band at 0.85 GHz. Figure 1 shows the typical geometry of one of a series of resonators under test. The device is a YBCO strip with thickness of about 450 nm and line width of 250 $\mu$m configured on a 0.5 mm-thick LAO substrate. [1] The patterned structure has YBCO only on the top LAO surface while the bottom surface of the substrate was glued to an aluminum microwave package housing that provides the necessary grounding. Prior work has explored the effects of nonuniform wet etching at the edges of the YBCO film associated with substrate twin-domain blocks (TDBs). Therefore, ion-milling lithography was used now to guarantee the edge evenness to better than 10 nm. For microwave characterization, the resonator was coupled through two capacitive gaps ($g$=200 $\mu$m) separating the input/output rf electrodes from the rf circuit delivering/measuring power $P_{IN}/P_{OUT}$ (as well as scattering parameters_ in the range from −50 to +10 dBm. For other measurements, both open ends of the HTS strip line were connected to a dc source (by contact pads I1 and I2) and voltage probes (V1, V2) for testing of superconducting and thermoelectric properties. For all the illustrated experiments (including the LTLSM characterization), the resonator was placed inside the vacuum cavity of a variable-temperature optical cryostat which stabilizes the temperature of the sample in the range 4–300 K with an accuracy of 1 mK.

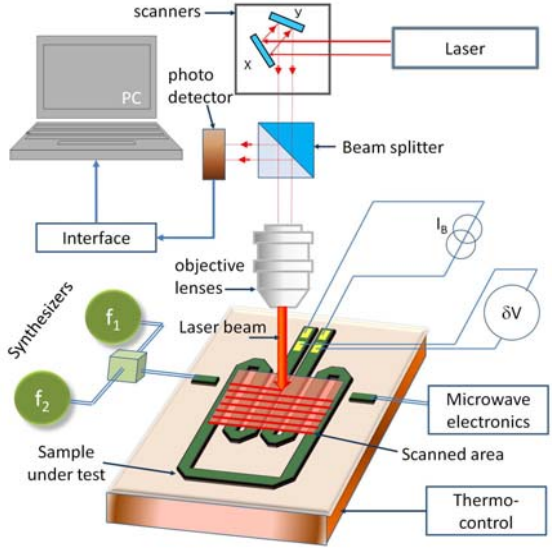

FIG. 2. Simplified schematic representation of the LSM experiment for simultaneous 2D visualization of optical, thermoelectric, dc, and microwave photoresponse of the tested resonator structure. Drawing is not to scale.

## III. METHODS

The method of LTLSM has been used as a key spatially-resolved instrument to probe local optical and superconducting properties, as well as the spatial variations in thermoelectric and electronic (both dc and rf) transport in YBCO/LAO superconducting microstrip resonators with micron-range resolution. The main capabilities, imaging modes, as well as the potential of the LTLSM technique for noncontact imaging of different properties of HTS materials and devices are described in a review article. [15]

### A. Technique

The LTLSM technique relates to active laser probing that directs illuminating flux (focused laser probe) to different portions of a sample under test to produce a photoresponse signal, PR. By moving the focal point of the probe in a regular step-by-step pattern across the surface of the sample one can record a two-dimensional (2D) array of PR$(x, y)$ the $(x-)$ row and $(y-)$ column indices representing its local properties as a function of $(x, y)$ sample coordinates. This array is stored in the memory of a computer as LSM contrast voltage $\delta V(x, y)$, building up a 2D image (brightness, color, amplitude, etc.) of the optical, electronic, rf, and other properties of the superconductor and the substrate. For example, the field of reflectivity factor $\Re(x, y)$ of the sample can be imaged by using the photoresponse of an elemental photo detector $\delta V(x, y)$ calibrated in units of $\Re(x, y)$.

Figure 2 is a simplified schematic representation of the generic LTLSM system. It consists essentially of *(i)* optomechanical, *(ii)* cryogenic, and *(iii)* measuring, controlling, and processing electronic modules. The experiment is performed in the following way: first, a collimated beam of laser diode output (maximum power: $P_{LAS}^{\max} = 1mW$, wavelength: $\lambda_{LAS}$=670 nm) is focused by microscope objective lenses on the surface of the resonator chip into a 1 $\mu$m Gaussian laser probe. The probe is TTL intensity modulated with a typical frequency $f_M$=100 kHz and scanned in a raster pattern with equal steps by using the $(x, y)$ deflection mirror unit of two galvos controlled by the computer. The reflected laser power is converted by a photo-detector to a voltage response signal $\delta V(x, y)$ and analyzed by the (synchronized to $f_M$) lock-in amplifier having output, $PR(x, y) \propto \Re(x, y)$ to image optically distinguishable defects in both the YBCO film and the LAO substrate. The absorbed part of laser energy heats the sample on the thermal healing length scale $l_T$ = (3–4) $\mu$m for bolometric probing of all the thermally dependent properties of the device. [15] The thermal healing length depends on the thermal conductivity and heat capacity of both the film and substrate, as well as the intensity modulation frequency $f_M$. [16, 17] For our YBCO/LAO system with a perfect thermal contact at the interface, the estimated magnitude of temperature oscillation in the film due to the laser probe is $\delta T \sim$ 1 K at $f_M$=100 kHz and $P_{LAS}^{\max} = 1mW$.

### B. Imaging modes

*1. Thermoelectric contrast*

Room temperature ($T_B \sim$ 300 K) local heating of the YBCO film causes an off-diagonal thermoelectric (Seebeck) effect related to the flow of thermal energy through the microstructure of misoriented layers of the HTS film. Even in the case of an electrically unbiased sample, the probe generates a temperature difference $T_Z$ between the film top ($z$ = 0) and the film bottom ($z=d_f$), exciting a longitudinal thermoelectric voltage PR$(x, y) \sim \delta V_S(x, y)$ between V1 and V2 (see Figs. 1 and 2) because of anisotropy of the thermoelectric properties of HTS films. In the case of predominantly orthogonal heat diffusion into the substrate, the PR*(x, y)* is proportional to the tilt angle α between the crystallographic *c* axis of the film and the normal to its surface: $\delta V_S(x,y) = \Delta T_Z \Delta S(l_T/d_f)\sin 2\alpha$. Here, $\Delta S = S_{ab} - S_c$ is the difference between the value of the thermopower $S_c$ along the crystallographic *c* axis and the thermopower $S_{ab}$ in the (*a*, *b*) plane. The thermoelectric imaging (TEI) mode of LSM contrast is used in this article to identify the twin-domain structure and the crystallographic grain orientation in different areas of the HTS film.

*2. dc voltage contrast*

The origin of the bolometric PR(*x*, *y*) in a dc current-biased HTS structure is mainly attributed to local modulation of the resistivity $\rho(T)$ and the sample bias current density, $J_B$, of YBCO by the heating effect of the oscillating laser probe. In the general case, it can be expressed as follows: [15]

$$\delta V(x,y) = \left[\begin{array}{c} J_B(x,y) \times \dfrac{d\rho(x,y)}{dT} \\ + \rho(x,y) \times \dfrac{dJ_B(x,y)}{dT} \end{array}\right] \times \Lambda \delta T(x,y), \quad (1)$$

where $\Lambda \sim l_T$ is the characteristic radius of the laser-beam-induced temperature spot. As evident from Eq. (1), normal ($d\rho/dT \cong const$, but very small) and superconducting ($d\rho/dT \equiv 0$) phases give a small contribution to PR(*x*, *y*) at fixed $J_B$. (Here we ignore the voltage contribution arising from the kinetic inductance photoresponse [18] because the fastest time scale in the problem $1/f_M$ is too slow to produce significant voltage.) Well defined $\delta V(x, y)$ arises mainly within the temperature interval of the superconducting transition $\Delta T_c$ near the critical temperature $T_c$ of the sample. This feature allows for probing of local values of $\Delta T_c$ and $T_c$ in HTS, and can be used to characterize their electronic superconducting properties. At temperatures below $T_c$, the voltage LSM response

$$\delta V(x,y) \propto (\partial J_c/\partial T) \cdot (E(x,y)/J_c) \cdot \Lambda \cdot \delta T$$

arises when $J_B \geq J_c$ and electric field *E(x, y)* appears underneath the laser probe. This case can be used to image the distribution of $J_c(x, y)$, which in turn is the main characteristic responsible for NL effects in HTSs.

*3. Microwave contrast*

Also, different microwave (rf) imaging LSM modes were created for noncontact investigation of the spatially inhomogeneous rf transport in passive HTS microwave devices. In this case, the thermally induced changes in the microwave transmittance $S_{21}(f)$ in the device were measured as LSM photoresponse that can be expressed as follows:

$$PR \propto \frac{\partial \|S_{21}(f)\|^2}{\partial T}\delta T \quad (2)$$

The separate impacts of individual linear rf components were obtained through the partial derivatives of $S_{21}(f)$ in the form [19,20]

$$PR \sim \delta\|S_{21}(f)\|^2 = \frac{1}{2}\left(\begin{array}{c} \dfrac{\|S_{21}(f)\|^2}{\partial f_0}\dfrac{\partial f_0}{\partial T} \\ + \dfrac{\|S_{21}(f)\|^2}{\partial(1/2Q)}\dfrac{\partial(1/2Q)}{\partial T} \\ + \dfrac{\|S_{21}(f)\|^2}{\partial \hat{S}_{21}^2}\dfrac{\partial \hat{S}_{21}^2}{\partial T} \end{array}\right)\delta T(x,y) \quad (3)$$

where the three items in the brackets in Eq. (3) symbolize inductive ($PR_X$), resistive ($PR_R$), and insertion loss ($PR_{IL}$) components of LSM PR, respectively, and $f_0$ is the resonant rf frequency, $Q$ is the quality factor, and $\hat{S}_{12}$ is the maximum of the transmission coefficient. Evidently, the $PR_X(x, y)$ originates from frequency $\delta f_0$ tuning due to the laser-probe-induced modulation of the HTS magnetic penetration depth and kinetic inductance. This effect is used to image a quantity proportional to the square of the local current density $J_{rf}(x, y)$. The remaining components $PR_R(x, y)$ and $PR_{IL}(x, y)$ together reflect changes in Ohmic dissipation produced by the laser probe and, therefore, were used to image spatial variations in the dissipation by a procedure that is described in detail in Refs. [19] and [20].

In the nonlinear rf imaging LSM mode, the method of third order intermodulation distortion (IMD) is used to probe changes in IMD transmitted power $P_{2f2-f1}$ or $P_{2f1-f2}$ as a function of position *(x, y)* of the laser beam perturbation on the sample. These signals result from nonlinear mixing of two rf excitations of the resonator at frequencies $f_1$ and $f_2$. For example, the change in IMD transmitted power $P_{2f1-f2}$ is given by [15, 20]

$$\frac{\delta P_{2f_1-f_2}}{P_{2f_1-f_2}} \sim \left\{\begin{array}{c} \dfrac{\delta\lambda}{\lambda} - \dfrac{\delta J_{IMD}}{J_{IMD}} \\ -\dfrac{\int \delta\lambda\, R_s\, J_{rf}^2\, dS + \int \delta R_s\, \lambda\, J_{rf}^2\, dS}{\int \lambda\, R_s\, J_{rf}^2\, dS} \end{array}\right\} \quad (4)$$

where $R_s$ is the surface resistance, $J_{IMD}$ is the nonlinearity current scale [21], and $\delta J_{IMD}$ is the change in nonlinearity current scale caused by the laser heating. Hence, the LSM IMD photoresponse is related to changes in the local nonlinearity current scale as well as changes in penetration depth and surface resistance at the site of the perturbation. Detailed interpretation of the IMD method and the results obtained with it will be done in Sec. IV.

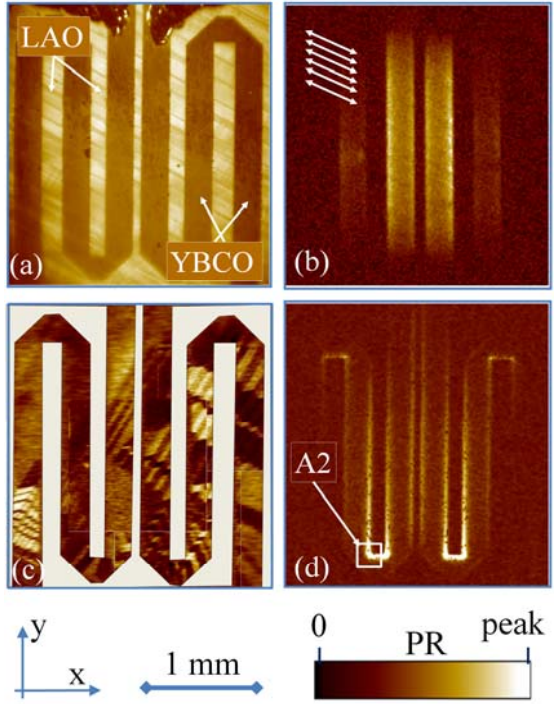

FIG. 3. Half-tone LSM images of (a) reflectivity map showing the twinned structure of the LAO substrate, (b) dc critical state indicating resistive regions of the YBCO strip at $T_B > T_c$ and $I_B < I_c$, (c) room temperature thermoelectric map illustrating the twin microstructure in YBCO, and (d) $\sim 3f_0$ standing wave pattern of $J_{rf}(x, y)$. All scan frames are 2.5x2.5 mm² acquired with a 12 $\mu$m laser probe. Bright areas indicate high LSM PR under laser-beam irradiation. Area A2 is imaged further in Fig. 9.

## IV. RESULTS

Figure 3 shows a set of four LSM images that were obtained in different imaging modes by probing the same large-scale 2.5x2.5 mm² area A1 (framed in Fig.1 by the dotted box) of the 1.85 GHz resonator. All these images are presented as 500x500 point arrays of data normalized to the maximum PR(x, y) in the range of each individual array. This allows clear visualization of the spatial distribution of different sample properties by using the same brightness contrast, where a white color is attributed to the maximum PR(x, y) while zero photoresponse is associated with the darkest brown color. As one can see from the optical reflectance LSM image [see Fig. 3(a)], the surface topography of the LAO substrate is dominated by an almost unidirectional strip-like twin-domain structure. The typical width of individual blocks varies between 10 and 150 $\mu$m. The fine interior microstructure of the LAO is not visible in the image due to the low spatial resolution of about 5 $\mu$m. In spite of this, no microcracks are observed in the YBCO films on the underlying domain boundaries (TDBs) of LAO, although the influence of these inhomogeneities on $J_c$ is clearly visible in Fig. 3(b). This LSM image shows the distribution of the resistive regions (bright areas) in the YBCO strip that is DC biased by $J_B = 0.2 \cdot 10^3 A/cm^2 \simeq J_c$ at $T_B = 89K \cong 0.97 T_c$. In this case, PR(x, y) is found only in those regions of the YBCO that are in the critical state, while the rest (dark areas) remains superconducting. It is not a surprise that a spatial modulation of PR(x, y) mimics the direction of TDBs indicated in Fig. 3(b) by the double-arrow lines. No more than a 5% reduction of the local $J_c(x,y)$ on TDBs results from a nanoscale LAO surface roughness, with a relief depth on the order of 4-5 nm [4]. Another feature in the distribution of PR(x, y) is even more interesting: an obvious gradient of critical parameters of YBCO can be seen through a reduction of the resistive photoresponse from the centre of the structure toward its periphery. By measuring dc current and temperature dependencies of LSM PR(x, y) we elucidated the reason for this. The gradient of critical parameters is mainly an experimental artifact due to irradiative heating of the resonator structure by room temperature objects outside of the optical cryostat. Specific boundary conditions of sample cooling give rise to overheating of the center part of the sample by 0.1 K relative to the substrate edges. Using this observation, we realize that the superconducting inhomogeneities of the YBCO are in fact very small, showing a spatial spread of $T_c(x,y) \leq 0.2K$ and $J_c(x,y) \leq 7\%$ down to $T_B \simeq 87K$. It was impossible to measure $J_c(x,y)$ on the periphery of the YBCO structure due to a strong Joule-heating effect at higher $J_c(x,y)$ below 87K.

Figure 3(c) shows a thermoelectric LSM PR image, $\delta V_S(x,y)$, obtained in the same 2.5x2.5 mm² area of the resonator at $T_B = 300K$. At first glance, there is no visible spatial correlation between this image (indicating the features of the microstructure of the YBCO film) and the images shown in Figs. 3(a) and 3(b). Also, it is difficult to see a large-scale correlation of all these images with the $J_{rf}(x, y)$ distribution at $T_B = 89K$ and microwave power $P_{rf} = 0 dBm$ shown in Fig. 3(d). Note that the LSM images of $J_{rf}(x, y)$ were extracted from the distribution of inductive (PR$_X$) relying on its square-law dependence on $J_{rf}(x, y)$. Additionally, a third harmonic resonant frequency $\approx 3f_0$ was excited to a obtain standing wave pattern $J(L) = J_0^* \sin(3\pi L / L_0)$ along the longitudinal direction $L$ on a resonating strip of length $L_0$ [Fig. 3(d)]. This resonant mode was chosen to excite a maximum rf current density in the region of minimum dc critical current in the structure. To clear up the question of whether a spatial correlation between different properties of YBCO and LAO exists, we conducted additional experiments with a micrometer-range LSM spatial resolution.

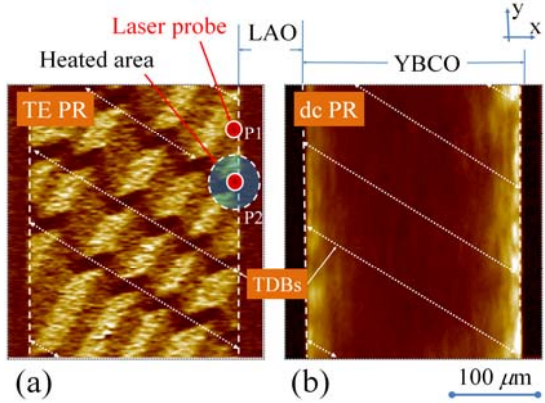
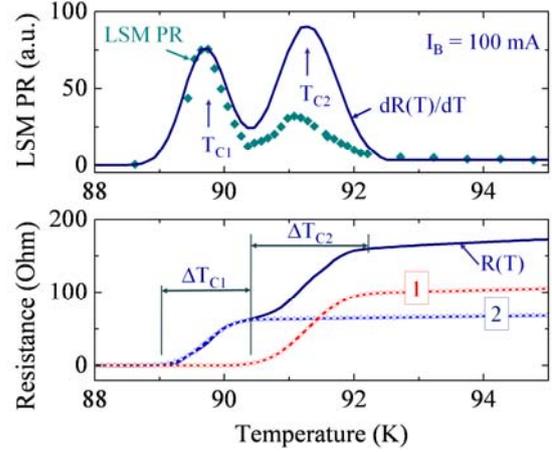

FIG. 4. The same 300x300 $\mu m^2$ area LSM scans of (a) twinned topology and (b) resistive dc transport properties of YBCO strip. Positions of the strip edges are outlined by the dashed lines. A regular array of lamellar (100)-type twin domains in YBCO is clearly visible in the thermoelectric LSM image (a). Brightness modulation between adjacent parquet-like structured twin-domains results from different orientations of individual twins inside domain relative to the direction of heat transport from the laser probe. Double-end-arrow dotted white lines show schematically the positions of kinks in the underlying LAO substrate. Points P1 and P2 indicate two positions of the stationary laser probe used to study the local electronic properties of the sample. The semitransparent area around P2 shows the region of thermal excitation outside the laser probe.

FIG. 5. Voltage LSM response (diamonds) at point P1 (a) and global (blue solid line) resistance R(T) (b) vs temperature of the HTS resonator at $I_B$=100 mA. In(a) the blue solid line is the temperature derivative of R(T) shown in (b), revealing peaks at transition temperatures $T_{c1}$ and $T_{c2}$. Curves "1" and "2" in (b) are reconstructions of R(T) that give rise to the global R(T) signal. In (b) the transition temperature widths are denoted as $\Delta T_{c1}$ and $\Delta T_{c2}$.

Figure 4(a) shows a $\delta V_S(x,y)$ LSM map of a 300x300 $\mu m^2$ area of the HTS resonator that was imaged at room temperature by a 1 $\mu m$ laser probe. The position of the 250 $\mu m$ wide YBCO strip is outlined by vertical dotted lines. As one can see, the twin structure of YBCO is formed in a regular array of lamellar (100)–type twins corrupted by LAO twin domain blocks (TDB). The position of the kinks coincides spatially with the directions of individual TDBs in LAO that was simultaneously imaged through the $\Re(x,y)$ LSM imaging mode, and is indicated by white arrows in Fig. 4. Surprisingly, the twin-domain microstructure of the YBCO film developed orthogonal to the TDBs in LAO. Moreover, the spacing of YBCO lamella is a few times smaller than that for the LAO TDBs. Figure 4(b) indicates that both types of (LAO and YBCO) twinning have an effect on the dc transport properties of the HTS strip. However, the influence of the LAO kinks suppresses the local $J_c(x,y)$, greatly increasing the dc resistive photoresponse exactly at the positions of the TDBs.

To give a quantitative estimation for the separate influence of twin/kink formation on local superconducting properties of YBCO, a stationary (not raster-scanned) LSM probe was used. In this case, a laser probe with a very small power (on the order of 1$\mu W$ producing local overheating $\delta T \leq 10^{-3} K$) is focused at a fixed position on the YBCO structure [see the red spot in Fig. 4(a)]. The photoresponse PR(x=const, y=const, F) is recorded as a function of some variable parameter F, such as $T_B$ or $I_B$. Figure 5(a) shows the temperature dependence of resistive LSM PR (diamonds) together with a scaled $dR(T_B)/dT$ curve (line) that was extracted from the "global" experimental dependence R(T) [blue line in Fig. 5(b)] of the HTS resonator. This data corresponds to the case when the focus of the laser probe is located at position P2 [see Fig. 4(a)] on the site of a single twin domain boundary.

As evident in Fig. 5, the peaks of locally measured LSM PR($T_B$) and globally measured $dR(T_B)/dT$ coincide in temperature. The only discrepancy is that the amplitudes of the LSM PR at $T_B = T_{C1}$ and at $T_B = T_{C2}$ are of different magnitudes compared to the $dR(T_B)/dT$ curve. Refocusing the laser probe at position (P2), where the twin lamella are not kinked, is accompanied by a decrease of the LSM PR at $T_B = T_{C1}$ and an increase at $T_B = T_{C2}$. This means that the temperature $T_B = T_{C1}$ corresponds to the midpoint of the superconducting transition of the kinks, while $T_B = T_{C2}$ is the midpoint of the transition of the twinned YBCO film without kinks. We believe that only the large-amplitude peak in the LSM PR($T_B$) curve corresponds to the property of the material underneath the laser probe. The appearance of the second smaller-amplitude peak is due to thermal modulation of the YBCO far from the focus, due to a thermal spot size [see dotted circle in Fig. 4(a)] of about 4.5 μm in diameter at the 100 kHz laser modulation rate. We also reconstructed the local values of $T_c$ and $\Delta T_c$ for different points of the HTS structure. An example of such a reconstruction is shown in Fig. 5(b) in curves 1 and 2. Note that the degradation of $T_c$ at the position of the kinks may be explained by sub-optimal oxygen doping resulting from the epitaxial growth of twin domain blocks.

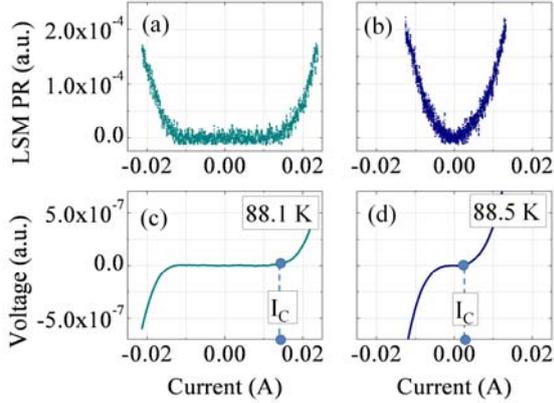

FIG. 6. Current dependencies of local dc LSM PR at 88.1 K (a) and 88.5 K (b) and the corresponding reconstructed IVCs [(c) and (d)] that were measured in position P1 [see Fig. 4(a)] of the YBCO strip. $I_c$ indicates the local value of the critical current.

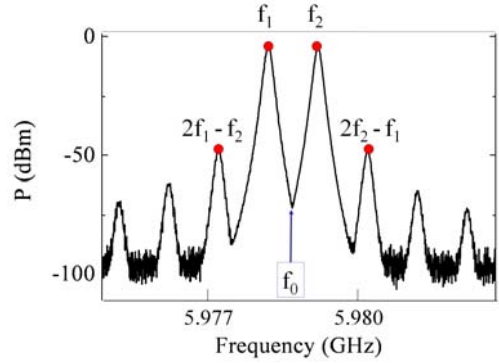

FIG. 7. The 1850 MHz resonator (Fig. 1) output spectrum at $T_B$=80 K generated as a result of the NL mixing of the two primary tones $P(f_1)$ = 0 dBm and $P(f_2)$ = 0 dBm at corresponding frequencies $f_1$ = 5.978 GHz and $f_2$ = 5.979 GHz, centered with $\Delta F$=1 MHz spacing around the third harmonic resonant frequency of the device, $f_0$ = 5.9785 GHz. The signals at frequencies $f_1$ and $f_2$ were used for the LSM imaging of both inductive and resistive components of rf PR, while the signals at $2f_1-f_2$ and $2f_2-f_1$ were used for imaging of NL sources. The signal at $f_0$ was used for LSM mapping to probe the spatial distribution of local insertion losses.

It is clear from the result above that the TDBs in the YBCO/LAO structure are the mostly limiting structural features for dc superconducting transport properties. This fact is directly supported by experiments on local probing of $J_c(x, y)$. Figures 6(a) and 6(b) show the $I_B$-dependence of LSM PR at the location of one kink [point P2 in the Fig. 4(a)] at two different temperatures $T_B$ close to $T_c$. By performing numerical integration of the data, it is possible to reconstruct the local current-voltage characteristics (IVCs) in the area of laser heating. Figures 6(c) and 6(d) show the corresponding IVCs calculated from the LSM PR data. A value of $J_C^{DC} \cong 10^9 \, A/m^2$ was obtained for individual TDBs in YBCO at 80 K from a temperature series of IVCs. Values of $J_c$ for the other areas of the structure were not measured due to large heating effects.

The method of stationary optical beam measurement has been applied for investigation of HF properties of the YBCO/LAO resonator at the same two locations (P1 and P2 in Fig. 4). We have shown earlier that the IMD spectrum of the resonating device (see Fig. 7) contains a full set of rf harmonics sufficient for LSM characterization of local inductive and resistive components of the linear response of the surface impedance, as well as its nonlinear (NL) contribution. In this case, two input tones (see Figs. 2 and 7) are applied at frequencies $f_1$ and $f_2$ straddling the device resonance frequency, $f_0$. A zero-frequency span mode of the spectrum analyzer is combined with a lock-in amplifier technique to separate the laser-probe-induced modulation of the rf output spectrum [$PR_{rf}(f=const)$] at fixed $f$, specified by the filled circles in Fig. 7. Changes in $PR_{rf}(f_1)$ and $PR_{rf}(f_2)$ at fixed positions $(x_i, y_j)$ of the laser beam perturbation on the sample were collected to probe the local current density $J_{rf}(x_i, y_j, P)$, and surface resistance $R_S(x_i, y_j, P)$ in compliance with the partition method described in Ref. [22], while change in $PR_{rf}(f_0)$ are used to probe insertion loss variations [20]. The IMD HF response $PR_{rf}(2f_1-f_2)$ either $PR_{rf}(2f_2-f_1)$ was utilized to probe local *microscopic* sources of NL response [15, 19].

Microwave power (with calibrated $J_{rf}$) dependencies of $PR_X(J_{rf})$ and $PR_R(J_{rf})$ at laser-beam position P1 are shown in Fig. 8. The procedure of LSM PR$(x, y)$ calibration to determine the absolute amplitude of $J_{rf}(x, y)$ is described in Ref. [15]. We find [see log-log plot in Figs. 8(a)] that a square-law dependence of $PR_X(J_{rf})$ on $J_{rf}$ is unchanged from zero amplitude of $J_{rf}$ up to $J_{rf}$ = 2.6x10$^9$ A/m$^2$, the maximum possible experimental rf current, limited by the leveled output of the synthesizer. One interesting feature is observed in the $PR_R(J_{rf})$ dependence seen in Fig. 8(b). The amplitude of the resistive component of rf LSM PR remains almost zero until a critical current indicated as $J_C^{rf}$ by a dotted line. Its value of about $J_C^{rf}$ = 1.16x10$^9$ A/m$^2$ at 80 K is in good agreement with $J_C^{dc} \cong 10^9 A/m^2$ measured at the same spatial point P2 and under the same conditions. Such agreement is also seen at point P1 of the sample where the twin domain structure of the YBCO film is free from TBDs. It should be noted that throughout this device we find a one-to-one correspondence between $J_C^{rf}(x,y)$, $J_C^{dc}(x,y)$, and $J_C^{IMD}(x,y)$, which is identified from the bend from cubic to square-law dependence of $P_{IMD}(P_{rf})$ with increasing rf power.

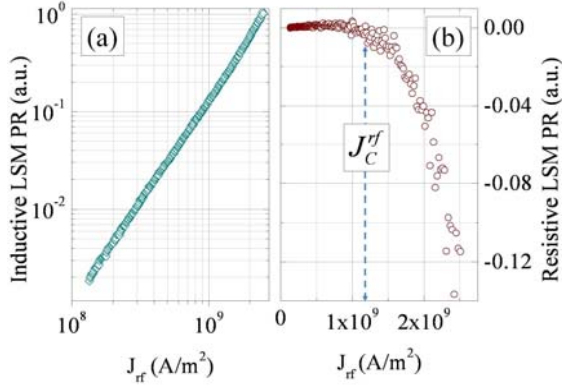

FIG. 8. Log-log plot of inductive LSM PR vs $J_{rf}$ showing classic square-law dependence (a) and linear plot of the $PR_R\_(J_{rf})$ dependence on $J_{rf}$ showing the local value of rf critical current density $J_C^{rf}$ at the point P1 at 80 K.

It has been shown by us earlier [19,20,22] that the sources of microwave nonlinearities may not be uniformly distributed across the film, and that IMD PR is localized near the peaks in $J_{rf}(x, y)$ on the inner corners of the patterned HTS structure. These areas of very large rf current densities are expected to be the principal candidates for generation of the most powerful local sources of NL component of rf response in the device, and were studied in detail by using different LSM imaging modes. As an example, a 50x50 μm² region is chosen for LSM scanning in the vicinity of one of the inner corners of the HTS patterned structure close to the point of maximum $J_{rf}(x, y)$ in the rf standing wave pattern of the resonator. [Frame A2 in Fig. 3(d)].

Fig. 9(a) shows 2-D LSM PR $\delta V_{dc}(x,y)$ image of this region at $J_B \cong J_c^{dc}$ below $T_c$. The YBCO/LAO patterned edge is roughly outlined by the white dashed line. Brighter regions in Fig. 9(a) correspond to higher amplitudes of the $\delta V_{dc}(x,y)$ associated with dc-Ohmic dissipation along the trajectory of magnetic vortex motion. This clearly indicates that vortices are predominantly nucleated at the corner and are channeled by the twin-domain structure of the YBCO film. The topology of this structure is quite different from that of the underlying LAO substrate, with the exception of the (kinked) areas indicated in the thermoelectric image [see Fig. 9(e)] by red arrows. The existence of such a modified YBCO twin structure is evidence for a partial (or even full) loss of epitaxial properties of the YBCO/LAO system during growth of a thick YBCO film. In spite of the fact that HF surface resistance is some analogue of dc resistivity at microwaves, there is no spatial correlation between these two electronic transport properties of YBCO film found by LSM imaging at all $J^{dc}(x,y) \cong J^{rf}(x,y)$. Figure 9(b) shows just a case of the $PR_R(x, y)$ distribution that has been LSM imaged keeping the same equal $J^{rf}(x,y)$ and $J^{dc}(x,y)$ for Figs. 9(b) and 9(a). As seen from Fig. 9(b), the distribution of $PR_R(x,$ $y$) is concentrated mainly at the YBCO strip edges, reaching local maxima near the inner corner where the rf current streamlines move closer together [see Fig. 9(d)]. The spatial modulation of $PR_R(x, y)$ along the edges is also clear visible in the framed white box that shows the LSM image with enhanced contrast (by ten times). The white arrows in Fig. 9(b) repeat those from Fig. 9(e) marking positions of inter-twin planes. These planes give a noticeable contribution to rf resistivity of YBCO, and cause IMD photoresponse in the same positions. [see Fig. 9(f)].

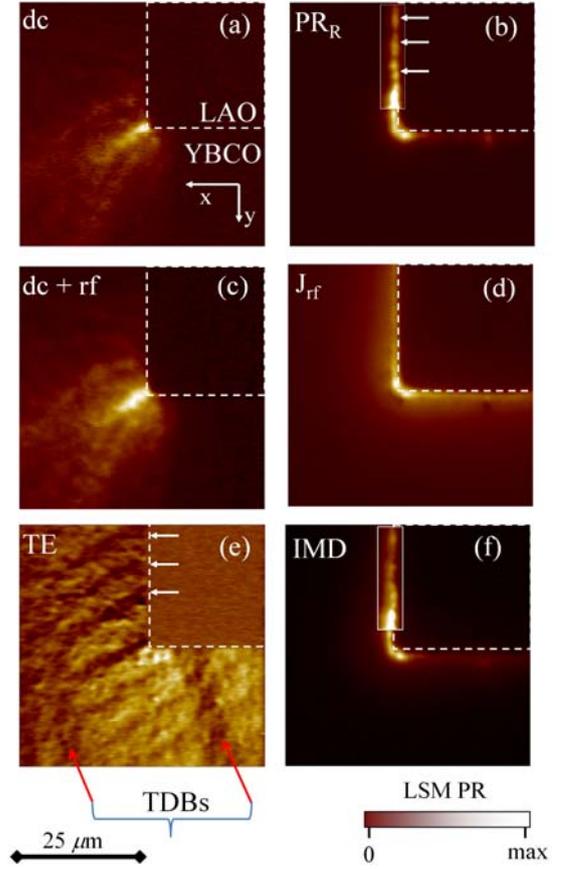

FIG. 9. Detail 50x50 μm² LSM images of dc dissipative regions (a), resistive rf response (b), dc photoresponse due to combined dc and rf currents (c), as well as distribution of $J_{rf}(x, y)$ (d), thermoelectric (e) and IMD (f) properties of YBCO microstrip at the position of one of the inner corners of the patterned structure that sharply changes the direction of dc and rf current flow (area A2 in Fig. 3). The position of YBCO/LAO patterned edge is outlined by the dashed white line. White arrows in (b) and (e) show the expected positions of inter-twin boundaries.

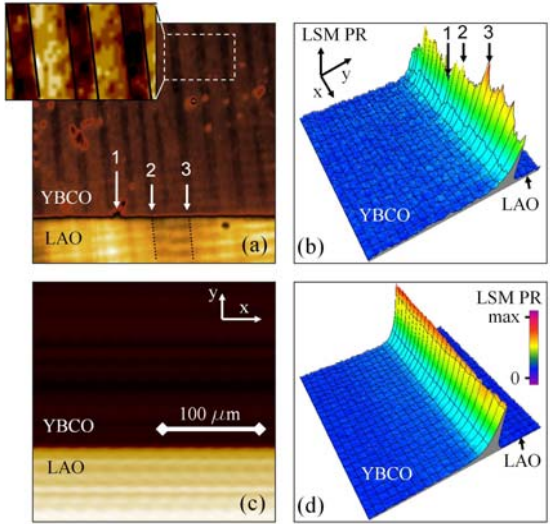

FIG. 10. 250x250 $\mu m^2$ reflective LSM images [(a) and (c)] and corresponding microwave PR images [(b) and (d)] showing the influence of twin-domain structure on rf transport properties of a thin YBCO film patterned edge. The inset in (a) shows the reflectivity of the YBCO in the area framed by the dashed box, and is replotted with higher contrast.

However, in contrast to $J^{dc}(x,y)$, the influence of both linear and nonlinear microwave transport does not spread deep inside the YBCO film, but always remains close to the strip edges. Surprisingly, the spatially modulated structure of LSM PR($x, y$) reverts to a TDB-dependent form for the case of simultaneously-applied dc and rf currents $J(x,y) = J^{dc}(x,y)/2 + J^{rf}(x,y)/2$, as depicted in Fig. 9(c). Here we substitute half of the dc current bias with an rf bias current so that the total current density is the same as in Fig. 9(a). Distinct trajectories of magnetic penetration along TDBs are visible again in this LSM image. However, the combination of simultaneous dc and rf current bias is different from both LSM distributions taken with $J^{rf}(x,y)$ and $J^{dc}(x,y)$ separately. In the case of rf excitation of the device in the presence of small dc bias, the rf field give rise to addition dissipation in the pre-existing vortex state of YBCO. The enhanced LSM dc PR in the presence of rf current may be due to depinning or edge-barrier reduction in the vortices in the presence of an ac bias.

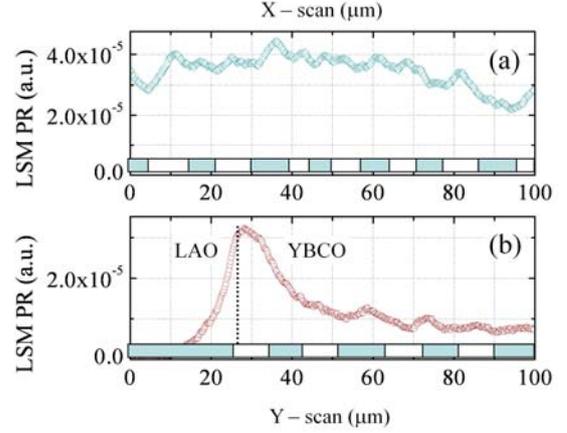

FIG. 11. Line-scan modulation of microwave LSM PR (a) along ($x$-scan) and (b) across ($y$-scan) the patterned YBCO/LAO edge strip in Fig. 10. The striped segments show the position of TDBs along the line scans. The influence of the twin-domain structure is clearly visible.

It is interesting to compare the results above with similar data obtained on thin-film HTS devices. Fig. 10(a) shows reflective LSM images collected in a 250x250 $\mu m^2$ area of a 230 nm thick YBCO strip patterned on an LAO substrate in the form of the resonator shown in Fig. 1. The complicated twin-domain structure of LAO has a strong influence on surface topology of YBCO. The fine zones [Fig. 10(a), inset] between the ~20 $\mu m$ wide strips are YBCO surface corrugations arising from the underlying twin domain blocks of LAO. Figure 10(b) shows that these defects initiate an inhomogeneous rf current flow along the strip edges and can produce NL rf response due to overcritical current densities. For TDBs oriented parallel to the patterned edge [see reflective LSM map in Fig. 10(c)], the distribution of $J^{rf}(x,y)$ is very smooth and uniform [see Fig. 10(d)]. In this case, the TDBs play the role of effective pinning planes preventing magnetic vortex entry into the film from the edges. The orthogonal orientation of the TBDs gives a significant weakening of the local edge barrier, as seen from the single-line-scan profile of LSM PR plotted in Fig. 11(a). The amplitude of this profile is qualitatively proportional to the $J^{rf}(x)$ from Fig. 10(b), where the $x$ direction coincides with the position of the YBCO/LAO interface. The striped bar below the profile is inserted to show the positions of individual TDBs along the LSM scan. As evident, the spatial modulation of $J^{rf}(x)$ along the edge varies by a factor of two under the influence of the TDBs, showing a proportional degradation of the superconducting properties of YBCO. The same degradation is clearly visible in the *Y*-scan imaged in Fig. 11(b) across the YBCO strip edge.

Analyzing the results presented in Figs. 10 and 11, one can come to the conclusion that epitaxial growth of thin YBCO films on LAO substrates with corrugated twin-domain surface topology causes unpredictable variations in rf superconducting transport properties of the HTS film depending on the orientation of the underlying TDB structure of the substrate. The best way to improve the rf properties of manufactured YBCO/LAO devices is to pattern their stripline geometry with an orientation of the strip edges precisely along the direction of the TDBs.

## V. CONCLUSION

In summary, we have developed a number of different imaging modes of LTLSM to probe local optical, superconducting, electronic (both dc and rf), as well thermoelectric transport properties of a YBCO/LAO superconducting microstrip resonator with a micron-range spatial resolution. Local sources of microwave NL were simultaneously mapped by using the LTLSM with two-tone rf IMD contrast as a function of $(x, y)$ position of the laser beam perturbation on the sample. The influence of the orientation of the individual twin-domain YBCO blocks on its NL properties was analyzed in detail. The results show the direct spatial correlation between microwave electronic transport properties of thin (< 0.5 $\mu$m thick) YBCO films and the underlying twin-domain topology of the LAO substrate. Thick (>2$\lambda_L$ thick) YBCO films are less subject to the effects of the LAO surface corrugations due to the loss of epitaxial morphology with the underlying LAO microstructure during the growth process. However, the YBCO twin-domain topology highly influences its dc and rf transport properties giving rise to formation of local sources of NL response. Despite the lack of spatial correlation between dc and rf current patterns found in thick YBCO films, their critical superconducting properties display the same scale of local critical current densities $J_c(x, y)$ measured with dc and rf LSM imaging modes at the same positions through the whole YBCO structure. In addition, the scale of local NL current densities $J_{IM}(x, y)$ in different areas of the YBCO microstrip also quantitatively coincides with the scale of local critical current densities $J_c(x, y)$ measured at the same positions.

## ACKNOWLEDGMENTS

The authors thank Monica Lilly for providing the AFM measurements of the resonator samples and D. W. Face for important information about the films. We acknowledge the support of the Fundamental Researches State Fund of Ukraine and German International Bureau of the Federal Ministry of Education and Research (BMBF) under Grant Project No. UKR08/011, a NASU program on "nanostructures, materials and technologies," the U.S. Department of Energy, and the Office of Naval Research/UMD AppEl Center, task D10.